%% file: main.tex
\documentclass[sigconf, noacm, dvipsnames]{acmart}

\AtBeginDocument{%
  \providecommand\BibTeX{{%
    \normalfont B\kern-0.5em{\scshape i\kern-0.25em b}\kern-0.8em\TeX}}}

\newcommand\ignore[1]{ }

\usepackage{xcolor}
\usepackage{multirow}

\definecolor{airforceblue}{rgb}{0.36, 0.54, 0.66}
\definecolor{dodgerblue}{rgb}{0.12, 0.56, 1.0}
\definecolor{brandeisblue}{rgb}{0.0, 0.44, 1.0}
\definecolor{brickred}{rgb}{0.8, 0.25, 0.33}
\definecolor{eggplant}{rgb}{0.38, 0.25, 0.32}
\definecolor{byzantium}{rgb}{0.44, 0.16, 0.39}

\newcommand{\juan}[1]{\textcolor{black}{#1}}
\newcommand{\juang}[1]{\textcolor{black}{#1}}

\definecolor{ddgreen}{rgb}{0.00, 0.50, 0.00}

\usepackage{tikz}
\usetikzlibrary{shapes.geometric, arrows, positioning, shadows}

\pgfdeclarelayer{background}
\pgfsetlayers{background,main}

\tikzstyle{texte} = [above]

\tikzstyle{stop} = [rectangle, rounded corners, minimum width=3cm, minimum height=1cm,text centered, draw=black, fill=violet!30, drop shadow]
\tikzstyle{io} = [trapezium, trapezium left angle=70, trapezium right angle=110, text centered, text width=5cm, draw=black, fill=blue!30, drop shadow]
\tikzstyle{processcpu} = [rectangle, text centered, text width=5cm, draw=black, fill=orange!30, drop shadow]
\tikzstyle{processdpu} = [rectangle, text centered, text width=5cm, draw=black, fill=red!30, drop shadow]
\tikzstyle{decision} = [diamond, aspect=2, text centered, draw=black, fill=green!30, drop shadow]
\tikzstyle{arrow} = [thick,->,>=stealth]

\usepackage{algorithm}
\usepackage{algpseudocode}
\usepackage{stmaryrd}
\usepackage{amsmath}
\usepackage{listings}
\usepackage{subcaption}

\definecolor{mygreen}{rgb}{0,0.6,0}
\definecolor{mygray}{rgb}{0.5,0.5,0.5}
\definecolor{mymauve}{rgb}{0.58,0,0.82}

\lstdefinestyle{myC}{
  language=Matlab,
  backgroundcolor=\color{white},  
  basicstyle=\footnotesize,        
  breakatwhitespace=false,  
  breaklines=true,       
  captionpos=b,                   
  commentstyle=\color{mygreen},    
  deletekeywords={...},           
  escapechar=\%,
  xleftmargin=0pt,
  xrightmargin=0pt,
  aboveskip=\medskipamount,
  belowskip=\medskipamount,
  extendedchars=true,            
  keepspaces=true,               
  keywordstyle=\color{blue},      
  language=C++,              
  morekeywords={__builtin_mul_sl_ul_rrr,__builtin_mul_sl_sh_rrr,mul_ul_ul,mul_sh_ul,mul_sh_sh,mul_sl_ul,mul_sl_sh,lbs,lhs,move,lsl_add,lw,add,sw,jneq,mem_alloc, *,...},          
  numbers=left,                   
  numbersep=1pt,                   
  numberstyle=\tiny\color{mygray}, 
  rulecolor=\color{black},     
  showspaces=false,                
  showstringspaces=false,          
  showtabs=false,                  
  stepnumber=1,                    
  stringstyle=\color{mymauve},     
  tabsize=2,	                   
  title=\lstname                   
}

\definecolor{bluehl}{rgb}{0.8,0.874,1}
\definecolor{pinkhl}{rgb}{0.992156863,0.847058824,1}
\definecolor{macaroniandcheese}{rgb}{1.0, 0.74, 0.53}
\definecolor{mossgreen}{rgb}{0.68, 0.87, 0.68}
\definecolor{greenhl}{rgb}{0.835,0.996,0.939}
\definecolor{yellowhl}{rgb}{0.996,0.957,0.8}
\definecolor{palecerulean}{rgb}{0.61, 0.77, 0.89}
\definecolor{gray(x11gray)}{rgb}{0.75, 0.75, 0.75}

\newcommand{\tsc}[1]{\textsuperscript{#1}} 
\newcommand{\affilETH}{\tsc{1}}
\newcommand{\affilUPM}{\tsc{2}}

\AtBeginDocument{\DeclareCaptionSubType{lstlisting}}

\setcopyright{none}
\renewcommand\footnotetextcopyrightpermission[1]{}

\makeatletter
\def\bstctlcite{\@ifnextchar[{\@bstctlcite}{\@bstctlcite[@auxout]}}
\def\@bstctlcite[#1]#2{\@bsphack
  \@for\@citeb:=#2\do{%
    \edef\@citeb{\expandafter\@firstofone\@citeb}%
    \if@filesw\immediate\write\csname #1\endcsname{\string\citation{\@citeb}}\fi}%
  \@esphack}
\makeatother

\begin{document}
\bstctlcite{IEEEexample:BSTcontrol}

\title{Machine Learning Training on \\ a Real Processing-in-Memory System}

\author{
 {%
     Juan Gómez-Luna$^1$\quad 
     Yuxin Guo$^1$\quad 
     Sylvan Brocard$^2$\quad 
     Julien Legriel$^2$
 }
}
\author{
 {
     Remy Cimadomo$^2$\quad
     Geraldo F. Oliveira$^1$\quad
     Gagandeep Singh$^1$\quad
     Onur Mutlu$^1$
 }
}


\affiliation{
\institution{
      \vspace{5pt}
      \affilETH ETH Zürich \quad
      \affilUPM UPMEM \quad
  }
}

\pagestyle{plain}

\begin{abstract}

\ignore{
Training machine learning algorithms is a computationally intensive process, which is frequently memory-bound due to repeatedly accessing large training datasets. 
As a result, processor-centric systems (e.g., CPU, GPU) suffer from costly data movement between memory units and processing units, which consumes large amounts of energy and execution cycles. 
Memory-centric computing systems, i.e., computing systems with processing-in-memory (PIM) capabilities, can alleviate this data movement bottleneck.

Our goal is to understand the potential of modern general-purpose PIM architectures to accelerate machine learning training. 
To do so, we (1) implement several representative classic machine learning algorithms (namely, linear regression, logistic regression, decision tree, K-means clustering) on a real-world general-purpose PIM architecture, (2) characterize them in terms of accuracy, performance and scaling, and (3) compare to their counterpart implementations on CPU and GPU. 
Our experimental evaluation on a memory-centric computing system with more than 2500 PIM cores shows that general-purpose PIM architectures can greatly accelerate memory-bound machine learning workloads, when the necessary operations and datatypes are natively supported by PIM hardware. 
For example, our PIM implementation of decision tree is $27\times$ faster than a state-of-the-art CPU version on an 8-core Intel Xeon, and $1.34\times$ faster than a state-of-the-art GPU version on an NVIDIA A100. Our K-means clustering on PIM is $2.8\times$ and $3.2\times$ than state-of-the-art CPU and GPU versions, respectively. 

To our knowledge, our work is the first one to evaluate training of machine learning algorithms on a real-world general-purpose PIM architecture. 
We conclude this paper with several key observations, takeaways, and recommendations that can enlighten users of machine learning workloads, programmers of PIM architectures, and hardware designers and architects of future memory-centric computing systems. }


\input{sections/01-introduction}

\end{abstract}


\keywords{machine learning, processing-in-memory, regression, classification, clustering, benchmarking}

\maketitle

\begin{acks}
We acknowledge the generous gifts provided by our industrial partners, including ASML, Facebook, Google, Huawei, Intel, Microsoft, and VMware.
We acknowledge support from the Semiconductor Research Corporation and the 
ETH Future Computing Laboratory.

This extended abstract appears as an invited paper at the 2022 IEEE Computer Society Annual Symposium on VLSI (ISVLSI).
It is a summary version of our recent work~\cite{gomez2022experimental}. 

\end{acks}

\balance
{
  \bstctlcite{IEEEexample:BSTcontrol} 
   \let\OLDthebibliography\thebibliography
  \renewcommand\thebibliography[1]{
    \OLDthebibliography{#1}
    \setlength{\parskip}{0pt}
    \setlength{\itemsep}{0pt}
  }
  \bibliographystyle{IEEEtran}
  \bibliography{references}
}




\end{document}

%% file: sections/01-introduction.tex

Machine learning (ML) algorithms~\cite{geron2019, alpaydin2020, goodfellow2016, mohri2018, shalev2014, raschka2019} have become ubiquitous in many fields of science and technology due to their ability to learn \juan{from} and 
\juan{improve with} experience with minimal human intervention. These algorithms train by updating their model parameters in an iterative manner to improve the overall prediction accuracy.  
However, training machine learning algorithms is a computationally intensive process, which requires large amounts of training data. 
Accessing training data in current processor-centric systems (e.g., CPU, GPU) implies costly data movement between memory and processors, which results in high energy consumption and a large percentage of the total execution cycles. 
This data movement can become the bottleneck of the training process, if there is not enough computation and locality to amortize its cost. 

One way to alleviate the cost of data movement is \emph{processing-in-memory} (\emph{PIM})~\cite{mutlu2019,mutlu2020modern, ghoseibm2019, seshadri2020indram, mutlu2019enabling}, a \juan{data-centric} computing paradigm that places processing elements near or inside the memory arrays. 
PIM has been explored for decades~\cite{stone1970logic, Kautz1969, shaw1981non, kogge1994, gokhale1995processing, patterson1997case, oskin1998active, kang1999flexram, Mai:2000:SMM:339647.339673,murphy2001characterization, Draper:2002:ADP:514191.514197,aga.hpca17,eckert2018neural,fujiki2019duality,kang.icassp14,seshadri.micro17,seshadri.arxiv16,Seshadri:2015:ANDOR,seshadri2013rowclone,angizi2019graphide,kim.hpca18,kim.hpca19,gao2020computedram,chang.hpca16,xin2020elp2im,li.micro17,deng.dac2018,hajinazarsimdram,rezaei2020nom,wang2020figaro,ali2019memory,li.dac16,angizi2018pima,angizi2018cmp,angizi2019dna,levy.microelec14,kvatinsky.tcasii14,shafiee2016isaac,kvatinsky.iccd11,kvatinsky.tvlsi14,gaillardon2016plim,bhattacharjee2017revamp,hamdioui2015memristor,xie2015fast,hamdioui2017myth,yu2018memristive,syncron,fernandez2020natsa,cali2020genasm,kim.bmc18,ahn.pei.isca15,ahn.tesseract.isca15,boroumand.asplos18,boroumand2019conda,singh2019napel,asghari-moghaddam.micro16,DBLP:conf/sigmod/BabarinsaI15,chi2016prime,farmahini2015nda,gao.pact15,DBLP:conf/hpca/GaoK16,gu.isca16,guo2014wondp,hashemi.isca16,cont-runahead,hsieh.isca16,kim.isca16,kim.sc17,DBLP:conf/IEEEpact/LeeSK15,liu-spaa17,morad.taco15,nai2017graphpim,pattnaik.pact16,pugsley2014ndc,zhang.hpdc14,zhu2013accelerating,DBLP:conf/isca/AkinFH15,gao2017tetris,drumond2017mondrian,dai2018graphh,zhang2018graphp,huang2020heterogeneous,zhuo2019graphq,santos2017operand,ghoseibm2019,wen2017rebooting,besta2021sisa,ferreira2021pluto,olgun2021quactrng,lloyd2015memory,elliott1999computational,zheng2016tcam,landgraf2021combining,rodrigues2016scattergather,lloyd2018dse,lloyd2017keyvalue,gokhale2015rearr,nair2015active,jacob2016compiling,sura2015data,nair2015evolution,balasubramonian2014near,xi2020memory,impica,boroumand2016pim,giannoula2022sparsep,giannoula2022sigmetrics,denzler2021casper,boroumand2021polynesia,boroumand2021icde,singh2021fpga,singh2021accelerating,herruzo2021enabling,yavits2021giraf,asgarifafnir,boroumand2021google_arxiv,boroumand2021google,amiraliphd,singh2020nero,seshadri.bookchapter17,diab2022high,diab2022hicomb,fujiki2018memory,zha2020hyper,mutlu.imw13,mutlu.superfri15,ahmed2019compiler,jain2018computing,ghiasi2022genstore,deoliveira2021IEEE,deoliveira2021,cho2020mcdram,shin2018mcdram,gu2020ipim,lavenier2020,Zois2018}. However, memory technology challenges prevented from its successful materialization in commercial products. For example, the limited number of metal layers in DRAM~\cite{weber2005current,peng2015design} makes conventional processor designs 
\juang{impractical} \juan{in commodity DRAM chips}~\cite{devaux2019,yuffe2011,christy2020,singh2017}.

\juan{Real-world} PIM systems have 
\juan{only recently been manufactured and commercialized}. 
The UPMEM company, for example, 
\juan{introduced} the first general-purpose \juan{commercial} PIM architecture~\cite{upmem,upmem2018,gomezluna2021benchmarking, gomezluna2022ieeeaccess, gomezluna2021cut}, which integrates small in-order cores near DRAM memory banks. 
High-bandwidth memory (HBM)-based HBM-PIM~\cite{kwon202125, lee2021hardware} and Acceleration DIMM (AxDIMM)~\cite{ke2021near} are Samsung's proposals that have been successfully tested 
\juan{via real} prototypes. HBM-PIM features \emph{Single Instruction Multiple Data} (\emph{SIMD}) units, which 
\juan{support} multiply-add and multiply-accumulate operations, near the banks in HBM layers~\cite{jedec.hbm.spec, lee.taco16}, and it is designed to accelerate neural network inference. AxDIMM is a near-rank solution that places an FPGA fabric on \juan{a DDR module} 
to accelerate specific workloads (e.g., recommendation inference). 
Accelerator-in-Memory (AiM)~\cite{lee2022isscc} is a GDDR6-based PIM architecture from SK Hynix with specialized units for multiply-accumulate and activation functions for deep learning. 
HB-PNM~\cite{niu2022isscc} is a 3D-stacked-based PIM architecture from Alibaba, which stacks a layer of LPDDR4 memory and a logic layer with specialized accelerators for \juan{recommendation} systems.

\begin{figure}[h]
\centering
\includegraphics[width=1.0\linewidth]{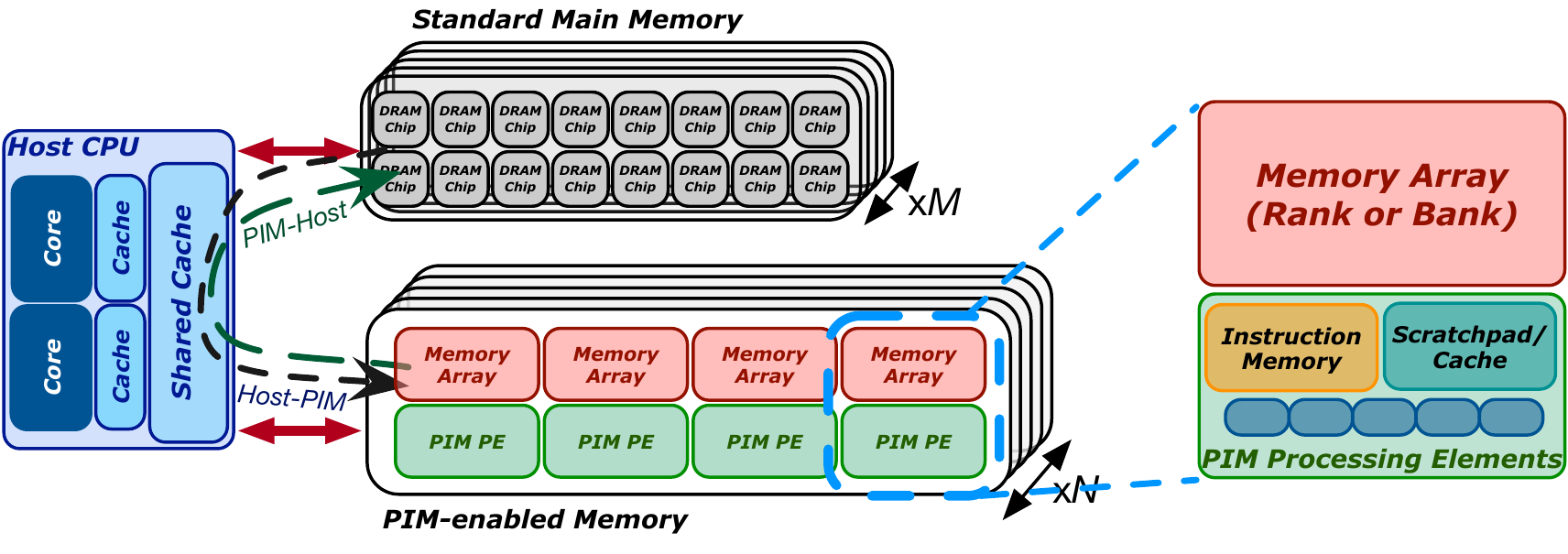}
\caption{High-level view of a state-of-the-art processing-in-memory system. \juan{The host CPU has access to $M$ standard memory modules and $N$ PIM-enabled memory modules.}}
\label{fig:scheme}
\end{figure}

\juan{These five real-world PIM systems} 
have some important \juan{common} characteristics, \juan{as depicted in Figure~\ref{fig:scheme}}. 
First, there is a host processor (CPU or GPU), typically with a deep cache hierarchy, which has access to (1) standard main memory, and (2) PIM-enabled memory (i.e., UPMEM DIMMs, HBM-PIM stacks, AxDIMM DIMMs, AiM GDDR6, HB-PNM LPDDR4). 
Second, the PIM-enabled memory \juan{chip} contains multiple PIM processing elements (PIM PEs), which have access to memory (either memory banks or ranks) with higher 
bandwidth \juan{and lower latency} than the host processor. 
Third, the \juan{PIM} processing elements (either \juan{general-purpose} cores, SIMD \juan{units, FPGAs, or specialized processors}) run at \juang{only} a few hundred megahertz, and have a small number of registers and 
\juan{relatively small} (or no) cache or scratchpad memory. 
Fourth, processing elements may not be able to communicate directly \juan{with each other} (e.g., UPMEM DPUs, HBM-PIM PCUs or AiM PUs in different chips), 
\juan{and} communication \juan{between them} happens via the host processor. 
Figure~\ref{fig:scheme} shows a high-level view of such a state-of-the-art processing-in-memory system. 

Our goal in this work is to quantify the potential of general-purpose PIM architectures for training of machine learning algorithms. 
To this end, we implement four representative \juan{classical} machine learning algorithms (linear regression~\cite{freedman2009statistical, yan2009linear}, logistic regression~\cite{freedman2009statistical, hosmer2013applied}, decision tree~\cite{suthaharan2016decision}, K-means clustering~\cite{Lloyd82leastsquares}) 
\juan{on} a general-purpose memory-centric system containing PIM-enabled memory, 
\juan{specifically} the UPMEM PIM architecture~\cite{upmem,upmem2018,gomezluna2021benchmarking, gomezluna2022ieeeaccess, gomezluna2021cut}. 
We do \emph{not} include training of deep learning 
algorithms in our study, since GPUs \juan{and TPUs} have a solid position as the preferred \juan{and highly optimized accelerators for 
deep learning} training~\cite{hwukirk2016.nn,chetlur2014cudnn,gao2017tetris,abadi2016tensorflow,runai_gpu,jouppi2017datacenter, jouppi2021ten}. 

Our PIM implementations of ML algorithms follow \juan{PIM} programming recommendations in recent literature~\cite{upmem-guide, upmem2018, gomezluna2021benchmarking, gomezluna2022ieeeaccess}. 
We apply several optimizations to overcome the limitations of existing general-purpose PIM architectures (e.g., limited instruction set, relatively simple pipeline, relatively low frequency) and take full advantage of the inherent strengths of PIM (e.g., large memory bandwidth, \juan{low memory latency}). 

We evaluate our PIM implementations in terms of \juang{training} accuracy, performance, and scaling characteristics on a \juan{real} memory-centric system with PIM-enabled memory~\cite{upmem, upmem-sdk, upmem-guide}. 
\juan{We run our experiments on a real-world PIM system~\cite{upmem} with 2,524 PIM cores running at 425 MHz, and 158 GB of DRAM memory.\footnote{The UPMEM-based PIM system has up to 2,560 PIM cores and 160 GB of DRAM.}}

Our experimental \juan{real system} evaluation provides 
\juan{new} observations and insights, 
\juan{including the following:} 
\begin{itemize}
\item ML training workloads that 
\juan{show memory-bound behavior} in processor-centric systems can greatly benefit from (1) fixed-point \juang{data} representation, (2) quantization~\cite{zmora2021quantization,gholamisurvey}, and (3) hybrid precision implementation~\cite{hopper, lee2022isscc} \juan{(without much accuracy loss) in PIM systems}, in order to 
\juan{alleviate} the lack of native support for floating-point 
and high-precision (i.e., 32- and 64-bit) \juang{arithmetic operations}. 
\item ML training workloads that require complex activation functions (e.g., sigmoid)~\cite{han1995influence} can take advantage of \emph{lookup tables} (\emph{LUTs})~\cite{ferreira2021pluto, deng2019lacc, gao2016draf} \juan{in PIM systems} instead of function approximation (e.g., Taylor series)~\cite{weisstein2004taylor}, when 
\juan{PIM systems} lack native support for those activation functions. 
\item 
\juan{Data can be placed and laid out such that} accesses of PIM cores to their nearby memory banks are 
streaming, 
\juan{which enables} better exploitation of the \juan{PIM} memory bandwidth. 
\item ML training workloads with large training datasets can greatly benefit from scaling \juan{the size of} PIM-enabled memory with PIM cores attached to memory banks. Training datasets can remain in memory without being moved to the host processor (e.g., CPU, GPU) in every iteration of the training process. Even if PIM cores need to communicate intermediate results via the host processor, this communication overhead is 
tolerable \juan{with proper overlap of computation and communication}. 
\end{itemize}



We compare our PIM implementations of linear regression, \juan{logistic} regression, decision tree, and K-means clustering to their state-of-the-art CPU and GPU counterparts. We observe that memory-centric systems with PIM-enabled memory can significantly outperform processor-centric systems for memory-bound ML training workloads, when the operations needed by the ML workloads are natively supported by PIM hardware (or can be replaced by \juan{efficient LUT implementations}). 

Our extended paper~\cite{gomez2022experimental} contains (1) detailed description of our PIM implementations of ML workloads; (2) comprehensive evaluation and \juang{comparisons to state-of-the-art CPU and GPU systems}; and (3) more insights about the suitability of ML workloads to the PIM system, programming recommendations for ML software developers, and suggestions and hints for future PIM architectures. 
We \juan{aim to} open-source all our PIM implementations of ML training workloads, training datasets, and evaluation scripts.

%% file: main.bbl
\begin{thebibliography}{100}
\providecommand{\url}[1]{#1}
\csname url@samestyle\endcsname
\providecommand{\newblock}{\relax}
\providecommand{\bibinfo}[2]{#2}
\providecommand{\BIBentrySTDinterwordspacing}{\spaceskip=0pt\relax}
\providecommand{\BIBentryALTinterwordstretchfactor}{4}
\providecommand{\BIBentryALTinterwordspacing}{\spaceskip=\fontdimen2\font plus
\BIBentryALTinterwordstretchfactor\fontdimen3\font minus
  \fontdimen4\font\relax}
\providecommand{\BIBforeignlanguage}[2]{{%
\expandafter\ifx\csname l@#1\endcsname\relax
\typeout{** WARNING: IEEEtran.bst: No hyphenation pattern has been}%
\typeout{** loaded for the language `#1'. Using the pattern for}%
\typeout{** the default language instead.}%
\else
\language=\csname l@#1\endcsname
\fi
#2}}
\providecommand{\BIBdecl}{\relax}
\BIBdecl

\bibitem{geron2019}
A.~G{\'e}ron, \emph{{Hands-on Machine Learning With Scikit-Learn, Keras, and
  TensorFlow: Concepts, tools, and techniques to build intelligent systems}},
  2019.

\bibitem{alpaydin2020}
E.~Alpaydin, \emph{{Introduction to Machine Learning}}, 2020.

\bibitem{goodfellow2016}
I.~Goodfellow \emph{et~al.}, \emph{{Deep Learning}}, 2016.

\bibitem{mohri2018}
M.~Mohri \emph{et~al.}, \emph{{Foundations of Machine Learning}}, 2018.

\bibitem{shalev2014}
S.~Shalev-Shwartz and S.~Ben-David, \emph{{Understanding Machine Learning: From
  Theory to Algorithms}}, 2014.

\bibitem{raschka2019}
S.~Raschka and V.~Mirjalili, \emph{{Python Machine Learning: Machine Learning
  and Deep Learning with Python, Scikit-Learn, and TensorFlow 2}}, 2019.

\bibitem{mutlu2019}
O.~Mutlu \emph{et~al.}, ``{Processing Data Where It Makes Sense: {E}nabling
  In-Memory Computation},'' \emph{MicPro}, 2019.

\bibitem{mutlu2020modern}
O.~Mutlu \emph{et~al.}, ``{A Modern Primer on Processing in Memory},''
  \emph{Emerging Computing: From Devices to Systems - Looking Beyond Moore and
  Von Neumann}, 2021, \url{https://arxiv.org/pdf/2012.03112.pdf}.

\bibitem{ghoseibm2019}
S.~Ghose \emph{et~al.}, ``{Processing-in-Memory: A Workload-Driven
  Perspective},'' \emph{IBM JRD}, 2019.

\bibitem{seshadri2020indram}
V.~Seshadri and O.~Mutlu, ``{In-DRAM Bulk Bitwise Execution Engine},''
  arXiv:1905.09822 [cs.AR], 2020.

\bibitem{mutlu2019enabling}
O.~Mutlu \emph{et~al.}, ``{Enabling Practical Processing in and near Memory for
  Data-Intensive Computing},'' in \emph{DAC}, 2019.

\bibitem{stone1970logic}
H.~S. Stone, ``{A Logic-in-Memory Computer},'' \emph{IEEE TC}, 1970.

\bibitem{Kautz1969}
W.~H. {Kautz}, ``{Cellular Logic-in-Memory Arrays},'' \emph{IEEE TC}, 1969.

\bibitem{shaw1981non}
D.~E. Shaw \emph{et~al.}, ``{The NON-VON Database Machine: A Brief Overview},''
  \emph{IEEE Database Eng. Bull.}, 1981.

\bibitem{kogge1994}
P.~M. Kogge, ``{EXECUBE - A New Architecture for Scaleable MPPs},'' in
  \emph{ICPP}, 1994.

\bibitem{gokhale1995processing}
M.~Gokhale \emph{et~al.}, ``{Processing in Memory: The Terasys Massively
  Parallel PIM Array},'' \emph{IEEE Computer}, 1995.

\bibitem{patterson1997case}
D.~Patterson \emph{et~al.}, ``{A Case for Intelligent RAM},'' \emph{IEEE
  Micro}, 1997.

\bibitem{oskin1998active}
M.~Oskin \emph{et~al.}, ``{Active Pages: {A} Computation Model for Intelligent
  Memory},'' in \emph{ISCA}, 1998.

\bibitem{kang1999flexram}
Y.~Kang \emph{et~al.}, ``{FlexRAM: Toward an Advanced Intelligent Memory
  System},'' in \emph{ICCD}, 1999.

\bibitem{Mai:2000:SMM:339647.339673}
K.~Mai \emph{et~al.}, ``{Smart Memories: A Modular Reconfigurable
  Architecture},'' in \emph{ISCA}, 2000.

\bibitem{murphy2001characterization}
R.~C. Murphy \emph{et~al.}, ``{The Characterization of Data Intensive Memory
  Workloads on Distributed PIM Systems},'' in \emph{Intelligent Memory
  Systems}.\hskip 1em plus 0.5em minus 0.4em\relax Springer.

\bibitem{Draper:2002:ADP:514191.514197}
J.~Draper \emph{et~al.}, ``{The Architecture of the DIVA Processing-in-Memory
  Chip},'' in \emph{SC}, 2002.

\bibitem{aga.hpca17}
S.~Aga \emph{et~al.}, ``{Compute Caches},'' in \emph{HPCA}, 2017.

\bibitem{eckert2018neural}
C.~Eckert \emph{et~al.}, ``{Neural Cache: Bit-serial In-cache Acceleration of
  Deep Neural Networks},'' in \emph{ISCA}, 2018.

\bibitem{fujiki2019duality}
D.~Fujiki \emph{et~al.}, ``{Duality Cache for Data Parallel Acceleration},'' in
  \emph{ISCA}, 2019.

\bibitem{kang.icassp14}
M.~Kang \emph{et~al.}, ``{An Energy-Efficient VLSI Architecture for Pattern
  Recognition via Deep Embedding of Computation in SRAM},'' in \emph{ICASSP},
  2014.

\bibitem{seshadri.micro17}
V.~Seshadri \emph{et~al.}, ``{Ambit: In-Memory Accelerator for Bulk Bitwise
  Operations Using Commodity DRAM Technology},'' in \emph{MICRO}, 2017.

\bibitem{seshadri.arxiv16}
V.~Seshadri \emph{et~al.}, ``{Buddy-RAM: Improving the Performance and
  Efficiency of Bulk Bitwise Operations Using DRAM},'' arXiv:1611.09988
  [cs:AR], 2016.

\bibitem{Seshadri:2015:ANDOR}
V.~Seshadri \emph{et~al.}, ``{Fast Bulk Bitwise AND and OR in DRAM},''
  \emph{CAL}, 2015.

\bibitem{seshadri2013rowclone}
V.~Seshadri \emph{et~al.}, ``{RowClone: Fast and Energy-Efficient In-DRAM Bulk
  Data Copy and Initialization},'' in \emph{MICRO}, 2013.

\bibitem{angizi2019graphide}
S.~Angizi and D.~Fan, ``{Graphide: A Graph Processing Accelerator Leveraging
  In-DRAM-computing},'' in \emph{GLSVLSI}, 2019.

\bibitem{kim.hpca18}
J.~Kim \emph{et~al.}, ``{The {DRAM} Latency {PUF}: Quickly Evaluating Physical
  Unclonable Functions by Exploiting the Latency--Reliability Tradeoff in
  Modern {DRAM} Devices},'' in \emph{HPCA}, 2018.

\bibitem{kim.hpca19}
J.~Kim \emph{et~al.}, ``{D-RaNGe: Using Commodity {DRAM} Devices to Generate
  True Random Numbers with Low Latency and High Throughput},'' in \emph{HPCA},
  2019.

\bibitem{gao2020computedram}
F.~Gao \emph{et~al.}, ``{ComputeDRAM: In-Memory Compute Using Off-the-Shelf
  DRAMs},'' in \emph{MICRO}, 2019.

\bibitem{chang.hpca16}
K.~K. Chang \emph{et~al.}, ``{Low-Cost Inter-Linked Subarrays (LISA): Enabling
  Fast Inter-Subarray Data Movement in DRAM},'' in \emph{HPCA}, 2016.

\bibitem{xin2020elp2im}
X.~Xin \emph{et~al.}, ``{ELP2IM: Efficient and Low Power Bitwise Operation
  Processing in DRAM},'' in \emph{HPCA}, 2020.

\bibitem{li.micro17}
S.~Li \emph{et~al.}, ``{DRISA: A DRAM-Based Reconfigurable In-Situ
  Accelerator},'' in \emph{MICRO}, 2017.

\bibitem{deng.dac2018}
Q.~Deng \emph{et~al.}, ``{DrAcc: A DRAM Based Accelerator for Accurate CNN
  Inference},'' in \emph{DAC}, 2018.

\bibitem{hajinazarsimdram}
N.~Hajinazar \emph{et~al.}, ``{SIMDRAM: A Framework for Bit-Serial SIMD
  Processing Using DRAM},'' in \emph{ASPLOS}, 2021.

\bibitem{rezaei2020nom}
S.~H.~S. {Rezaei} \emph{et~al.}, ``{NoM: Network-on-Memory for Inter-Bank Data
  Transfer in Highly-Banked Memories},'' \emph{CAL}, 2020.

\bibitem{wang2020figaro}
Y.~Wang \emph{et~al.}, ``{FIGARO: Improving System Performance via Fine-Grained
  In-DRAM Data Relocation and Caching},'' in \emph{MICRO}, 2020.

\bibitem{ali2019memory}
M.~F. Ali \emph{et~al.}, ``{In-Memory Low-Cost Bit-Serial Addition Using
  Commodity DRAM Technology},'' in \emph{{TCAS-I}}, 2019.

\bibitem{li.dac16}
S.~Li \emph{et~al.}, ``{Pinatubo: A Processing-in-Memory Architecture for Bulk
  Bitwise Operations in Emerging Non-Volatile Memories},'' in \emph{DAC}, 2016.

\bibitem{angizi2018pima}
S.~Angizi \emph{et~al.}, ``{PIMA-Logic: A Novel Processing-in-Memory
  Architecture for Highly Flexible and Energy-efficient Logic Computation},''
  in \emph{DAC}, 2018.

\bibitem{angizi2018cmp}
S.~Angizi \emph{et~al.}, ``{CMP-PIM: An Energy-efficient Comparator-based
  Processing-in-Memory Neural Network Accelerator},'' in \emph{DAC}, 2018.

\bibitem{angizi2019dna}
S.~Angizi \emph{et~al.}, ``{AlignS: A Processing-in-Memory Accelerator for DNA
  Short Read Alignment Leveraging SOT-MRAM},'' in \emph{DAC}, 2019.

\bibitem{levy.microelec14}
Y.~Levy \emph{et~al.}, ``{Logic Operations in Memory Using a Memristive Akers
  Array},'' \emph{Microelectronics Journal}, 2014.

\bibitem{kvatinsky.tcasii14}
S.~Kvatinsky \emph{et~al.}, ``{MAGIC---Memristor-Aided Logic},'' \emph{IEEE
  TCAS II: Express Briefs}, 2014.

\bibitem{shafiee2016isaac}
A.~Shafiee \emph{et~al.}, ``{ISAAC: A Convolutional Neural Network Accelerator
  with In-situ Analog Arithmetic in Crossbars},'' in \emph{ISCA}, 2016.

\bibitem{kvatinsky.iccd11}
S.~Kvatinsky \emph{et~al.}, ``{Memristor-Based IMPLY Logic Design Procedure},''
  in \emph{ICCD}, 2011.

\bibitem{kvatinsky.tvlsi14}
S.~Kvatinsky \emph{et~al.}, ``{Memristor-Based Material Implication (IMPLY)
  Logic: Design Principles and Methodologies},'' \emph{TVLSI}, 2014.

\bibitem{gaillardon2016plim}
P.-E. Gaillardon \emph{et~al.}, ``{The Programmable Logic-in-Memory (PLiM)
  Computer},'' in \emph{DATE}, 2016.

\bibitem{bhattacharjee2017revamp}
D.~Bhattacharjee \emph{et~al.}, ``{ReVAMP: ReRAM based VLIW Architecture for
  In-memory Computing},'' in \emph{DATE}, 2017.

\bibitem{hamdioui2015memristor}
S.~Hamdioui \emph{et~al.}, ``{Memristor Based Computation-in-Memory
  Architecture for Data-intensive Applications},'' in \emph{DATE}, 2015.

\bibitem{xie2015fast}
L.~Xie \emph{et~al.}, ``{Fast Boolean Logic Papped on Memristor Crossbar},'' in
  \emph{ICCD}, 2015.

\bibitem{hamdioui2017myth}
S.~Hamdioui \emph{et~al.}, ``{Memristor for Computing: Myth or Reality?}'' in
  \emph{DATE}, 2017.

\bibitem{yu2018memristive}
J.~Yu \emph{et~al.}, ``{Memristive Devices for Computation-in-Memory},'' in
  \emph{DATE}, 2018.

\bibitem{syncron}
C.~Giannoula \emph{et~al.}, ``{SynCron: Efficient Synchronization Support for
  Near-Data-Processing Architectures},'' in \emph{HPCA}, 2021.

\bibitem{fernandez2020natsa}
I.~Fernandez \emph{et~al.}, ``{NATSA: A Near-Data Processing Accelerator for
  Time Series Analysis},'' in \emph{ICCD}, 2020.

\bibitem{cali2020genasm}
D.~S. Cali \emph{et~al.}, ``{GenASM: A High-Performance, Low-Power Approximate
  String Matching Acceleration Framework for Genome Sequence Analysis},'' in
  \emph{MICRO}, 2020.

\bibitem{kim.bmc18}
J.~S. Kim \emph{et~al.}, ``{GRIM-Filter: Fast Seed Location Filtering in DNA
  Read Mapping Using Processing-in-Memory Technologies},'' \emph{BMC Genomics},
  2018.

\bibitem{ahn.pei.isca15}
J.~Ahn \emph{et~al.}, ``{PIM-Enabled Instructions: A Low-Overhead,
  Locality-Aware Processing-in-Memory Architecture},'' in \emph{ISCA}, 2015.

\bibitem{ahn.tesseract.isca15}
J.~Ahn \emph{et~al.}, ``{A Scalable Processing-in-Memory Accelerator for
  Parallel Graph Processing},'' in \emph{ISCA}, 2015.

\bibitem{boroumand.asplos18}
A.~Boroumand \emph{et~al.}, ``{Google Workloads for Consumer Devices:
  Mitigating Data Movement Bottlenecks},'' in \emph{ASPLOS}, 2018.

\bibitem{boroumand2019conda}
A.~Boroumand \emph{et~al.}, ``{CoNDA: Efficient Cache Coherence Support for
  near-Data Accelerators},'' in \emph{ISCA}, 2019.

\bibitem{singh2019napel}
G.~Singh \emph{et~al.}, ``{NAPEL: Near-memory Computing Application Performance
  Prediction via Ensemble Learning},'' in \emph{DAC}, 2019.

\bibitem{asghari-moghaddam.micro16}
H.~Asghari-Moghaddam \emph{et~al.}, ``{Chameleon: Versatile and Practical
  Near-DRAM Acceleration Architecture for Large Memory Systems},'' in
  \emph{MICRO}, 2016.

\bibitem{DBLP:conf/sigmod/BabarinsaI15}
O.~O. Babarinsa and S.~Idreos, ``{JAFAR: Near-Data Processing for Databases},''
  in \emph{SIGMOD}, 2015.

\bibitem{chi2016prime}
P.~Chi \emph{et~al.}, ``{PRIME: A Novel Processing-In-Memory Architecture for
  Neural Network Computation In ReRAM-Based Main Memory},'' in \emph{ISCA},
  2016.

\bibitem{farmahini2015nda}
A.~Farmahini-Farahani \emph{et~al.}, ``{NDA: Near-DRAM acceleration
  architecture leveraging commodity DRAM devices and standard memory
  modules},'' in \emph{HPCA}, 2015.

\bibitem{gao.pact15}
M.~Gao \emph{et~al.}, ``{Practical Near-Data Processing for In-Memory Analytics
  Frameworks},'' in \emph{PACT}, 2015.

\bibitem{DBLP:conf/hpca/GaoK16}
M.~Gao and C.~Kozyrakis, ``{HRL: Efficient and Flexible Reconfigurable Logic
  for Near-Data Processing},'' in \emph{HPCA}, 2016.

\bibitem{gu.isca16}
B.~Gu \emph{et~al.}, ``{Biscuit: {A} Framework for Near-Data Processing of Big
  Data Workloads},'' in \emph{ISCA}, 2016.

\bibitem{guo2014wondp}
Q.~Guo \emph{et~al.}, ``{3D-Stacked Memory-Side Acceleration: Accelerator and
  System Design},'' in \emph{WoNDP}, 2014.

\bibitem{hashemi.isca16}
M.~Hashemi \emph{et~al.}, ``{Accelerating Dependent Cache Misses with an
  Enhanced Memory Controller},'' in \emph{ISCA}, 2016.

\bibitem{cont-runahead}
M.~Hashemi \emph{et~al.}, ``{Continuous Runahead: Transparent Hardware
  Acceleration for Memory Intensive Workloads},'' in \emph{MICRO}, 2016.

\bibitem{hsieh.isca16}
K.~Hsieh \emph{et~al.}, ``{Transparent Offloading and Mapping (TOM): Enabling
  Programmer-Transparent Near-Data Processing in GPU Systems},'' in
  \emph{ISCA}, 2016.

\bibitem{kim.isca16}
D.~Kim \emph{et~al.}, ``{Neurocube: {A} Programmable Digital Neuromorphic
  Architecture with High-Density {3D} Memory},'' in \emph{ISCA}, 2016.

\bibitem{kim.sc17}
G.~Kim \emph{et~al.}, ``{Toward Standardized Near-Data Processing with
  Unrestricted Data Placement for GPUs},'' in \emph{SC}, 2017.

\bibitem{DBLP:conf/IEEEpact/LeeSK15}
J.~H. Lee \emph{et~al.}, ``{BSSync: Processing Near Memory for Machine Learning
  Workloads with Bounded Staleness Consistency Models},'' in \emph{PACT}, 2015.

\bibitem{liu-spaa17}
Z.~Liu \emph{et~al.}, ``{Concurrent Data Structures for Near-Memory
  Computing},'' in \emph{SPAA}, 2017.

\bibitem{morad.taco15}
A.~Morad \emph{et~al.}, ``{GP-SIMD Processing-in-Memory},'' \emph{ACM TACO},
  2015.

\bibitem{nai2017graphpim}
L.~Nai \emph{et~al.}, ``{GraphPIM: Enabling Instruction-Level PIM Offloading in
  Graph Computing Frameworks},'' in \emph{HPCA}, 2017.

\bibitem{pattnaik.pact16}
A.~Pattnaik \emph{et~al.}, ``{Scheduling Techniques for GPU Architectures with
  Processing-in-Memory Capabilities},'' in \emph{PACT}, 2016.

\bibitem{pugsley2014ndc}
S.~H. Pugsley \emph{et~al.}, ``{{NDC: Analyzing the Impact of 3D-Stacked
  Memory+Logic Devices on MapReduce Workloads}},'' in \emph{ISPASS}, 2014.

\bibitem{zhang.hpdc14}
D.~P. Zhang \emph{et~al.}, ``{TOP-PIM: Throughput-Oriented Programmable
  Processing in Memory},'' in \emph{HPDC}, 2014.

\bibitem{zhu2013accelerating}
Q.~Zhu \emph{et~al.}, ``{Accelerating Sparse Matrix-Matrix Multiplication with
  3D-Stacked Logic-in-Memory Hardware},'' in \emph{HPEC}, 2013.

\bibitem{DBLP:conf/isca/AkinFH15}
B.~Akin \emph{et~al.}, ``{Data Reorganization in Memory Using {3D}-Stacked
  {DRAM}},'' in \emph{ISCA}, 2015.

\bibitem{gao2017tetris}
M.~Gao \emph{et~al.}, ``{Tetris: Scalable and Efficient Neural Network
  Acceleration with 3D Memory},'' in \emph{ASPLOS}, 2017.

\bibitem{drumond2017mondrian}
M.~Drumond \emph{et~al.}, ``{The Mondrian Data Engine},'' in \emph{ISCA}, 2017.

\bibitem{dai2018graphh}
G.~Dai \emph{et~al.}, ``{GraphH: A Processing-in-Memory Architecture for
  Large-scale Graph Processing},'' \emph{IEEE TCAD}, 2018.

\bibitem{zhang2018graphp}
M.~Zhang \emph{et~al.}, ``{GraphP: Reducing Communication for PIM-based Graph
  Processing with Efficient Data Partition},'' in \emph{HPCA}, 2018.

\bibitem{huang2020heterogeneous}
Y.~Huang \emph{et~al.}, ``{A Heterogeneous PIM Hardware-Software Co-Design for
  Energy-Efficient Graph Processing},'' in \emph{IPDPS}, 2020.

\bibitem{zhuo2019graphq}
Y.~Zhuo \emph{et~al.}, ``{GraphQ: Scalable PIM-based Graph Processing},'' in
  \emph{MICRO}, 2019.

\bibitem{santos2017operand}
P.~C. Santos \emph{et~al.}, ``{Operand Size Reconfiguration for Big Data
  Processing in Memory},'' in \emph{DATE}, 2017.

\bibitem{wen2017rebooting}
W.-M. Hwu \emph{et~al.}, ``{Rebooting the Data Access Hierarchy of Computing
  Systems},'' in \emph{ICRC}, 2017.

\bibitem{besta2021sisa}
M.~Besta \emph{et~al.}, ``{SISA: Set-Centric Instruction Set Architecture for
  Graph Mining on Processing-in-Memory Systems},'' in \emph{MICRO}, 2021.

\bibitem{ferreira2021pluto}
J.~D. Ferreira \emph{et~al.}, ``{pLUTo: In-DRAM Lookup Tables to Enable
  Massively Parallel General-Purpose Computation},'' \emph{arXiv:2104.07699
  [cs.AR]}, 2021.

\bibitem{olgun2021quactrng}
A.~Olgun \emph{et~al.}, ``{QUAC-TRNG: High-Throughput True Random Number
  Generation Using Quadruple Row Activation in Commodity DRAMs},'' in
  \emph{ISCA}, 2021.

\bibitem{lloyd2015memory}
S.~Lloyd and M.~Gokhale, ``{In-memory Data Rearrangement for Irregular,
  Data-intensive Computing},'' \emph{Computer}, 2015.

\bibitem{elliott1999computational}
D.~G. Elliott \emph{et~al.}, ``{Computational RAM: Implementing Processors in
  Memory},'' \emph{IEEE Design \& Test of Computers}, 1999.

\bibitem{zheng2016tcam}
L.~Zheng \emph{et~al.}, ``{RRAM-based TCAMs for pattern search},'' in
  \emph{ISCAS}, 2016.

\bibitem{landgraf2021combining}
J.~Landgraf \emph{et~al.}, ``{Combining Emulation and Simulation to Evaluate a
  Near Memory Key/Value Lookup Accelerator},'' 2021.

\bibitem{rodrigues2016scattergather}
A.~Rodrigues \emph{et~al.}, ``{Towards a Scatter-Gather Architecture: Hardware
  and Software Issues},'' in \emph{MEMSYS}, 2019.

\bibitem{lloyd2018dse}
S.~Lloyd and M.~Gokhale, ``{Design Space Exploration of Near Memory
  Accelerators},'' in \emph{MEMSYS}, 2018.

\bibitem{lloyd2017keyvalue}
S.~Lloyd and M.~Gokhale, ``{Near Memory Key/Value Lookup Acceleration},'' in
  \emph{MEMSYS}, 2017.

\bibitem{gokhale2015rearr}
M.~Gokhale \emph{et~al.}, ``{Near Memory Data Structure Rearrangement},'' in
  \emph{MEMSYS}, 2015.

\bibitem{nair2015active}
R.~Nair \emph{et~al.}, ``{Active Memory Cube: A Processing-in-Memory
  Architecture for Exascale Systems},'' \emph{IBM JRD}, 2015.

\bibitem{jacob2016compiling}
A.~C. Jacob \emph{et~al.}, ``{Compiling for the Active Memory Cube},'' Tech.
  rep. RC25644 (WAT1612-008). IBM Research Division, Tech. Rep., 2016.

\bibitem{sura2015data}
Z.~Sura \emph{et~al.}, ``{Data Access Optimization in a Processing-in-Memory
  System},'' in \emph{CF}, 2015.

\bibitem{nair2015evolution}
R.~Nair, ``{Evolution of Memory Architecture},'' \emph{Proceedings of the
  IEEE}, 2015.

\bibitem{balasubramonian2014near}
R.~Balasubramonian \emph{et~al.}, ``{Near-Data Processing: Insights from a
  MICRO-46 Workshop},'' \emph{IEEE Micro}, 2014.

\bibitem{xi2020memory}
Y.~Xi \emph{et~al.}, ``{In-Memory Learning With Analog Resistive Switching
  Memory: A Review and Perspective},'' \emph{Proceedings of the IEEE}, 2020.

\bibitem{impica}
K.~Hsieh \emph{et~al.}, ``{Accelerating Pointer Chasing in 3D-Stacked Memory:
  Challenges, Mechanisms, Evaluation},'' in \emph{ICCD}, 2016.

\bibitem{boroumand2016pim}
A.~Boroumand \emph{et~al.}, ``{LazyPIM: An Efficient Cache Coherence Mechanism
  for Processing-in-Memory},'' \emph{CAL}, 2016.

\bibitem{giannoula2022sparsep}
C.~Giannoula \emph{et~al.}, ``{SparseP: Towards Efficient Sparse Matrix Vector
  Multiplication on Real Processing-In-Memory Systems},'' \emph{arXiv preprint
  arXiv:2201.05072}, 2022.

\bibitem{giannoula2022sigmetrics}
C.~Giannoula \emph{et~al.}, ``{Towards Efficient Sparse Matrix Vector
  Multiplication on Real Processing-in-Memory Architectures},'' in
  \emph{SIGMETRICS}, 2022.

\bibitem{denzler2021casper}
A.~Denzler \emph{et~al.}, ``Casper: Accelerating stencil computation using
  near-cache processing,'' \emph{arXiv preprint arXiv:2112.14216}, 2021.

\bibitem{boroumand2021polynesia}
A.~Boroumand \emph{et~al.}, ``{Polynesia: Enabling Effective Hybrid
  Transactional/Analytical Databases with Specialized Hardware/Software
  Co-Design},'' arXiv:2103.00798 [cs.AR], 2021.

\bibitem{boroumand2021icde}
A.~Boroumand \emph{et~al.}, ``{Polynesia: Enabling Effective Hybrid
  Transactional Analytical Databases with Specialized Hardware Software
  Co-Design},'' in \emph{ICDE}, 2022.

\bibitem{singh2021fpga}
G.~Singh \emph{et~al.}, ``{FPGA-based Near-Memory Acceleration of Modern
  Data-Intensive Applications},'' \emph{IEEE Micro}, 2021.

\bibitem{singh2021accelerating}
G.~Singh \emph{et~al.}, ``{Accelerating Weather Prediction using Near-Memory
  Reconfigurable Fabric},'' \emph{ACM TRETS}, 2021.

\bibitem{herruzo2021enabling}
J.~M. Herruzo \emph{et~al.}, ``{Enabling Fast and Energy-Efficient FM-Index
  Exact Matching Using Processing-Near-Memory},'' \emph{The Journal of
  Supercomputing}, 2021.

\bibitem{yavits2021giraf}
L.~Yavits \emph{et~al.}, ``{GIRAF: General Purpose In-Storage Resistive
  Associative Framework},'' \emph{IEEE TPDS}, 2021.

\bibitem{asgarifafnir}
B.~Asgari \emph{et~al.}, ``{FAFNIR: Accelerating Sparse Gathering by Using
  Efficient Near-Memory Intelligent Reduction},'' in \emph{HPCA}, 2021.

\bibitem{boroumand2021google_arxiv}
A.~Boroumand \emph{et~al.}, ``{Google Neural Network Models for Edge Devices:
  Analyzing and Mitigating Machine Learning Inference Bottlenecks},''
  \emph{arXiv preprint arXiv:2109.14320}, 2021.

\bibitem{boroumand2021google}
A.~Boroumand \emph{et~al.}, ``{Google Neural Network Models for Edge Devices:
  Analyzing and Mitigating Machine Learning Inference Bottlenecks},'' in
  \emph{PACT}, 2021.

\bibitem{amiraliphd}
A.~Boroumand, ``{Practical Mechanisms for Reducing Processor-Memory Data
  Movement in Modern Workloads},'' Ph.D. dissertation, Carnegie Mellon
  University, 2020.

\bibitem{singh2020nero}
G.~Singh \emph{et~al.}, ``{NERO: A Near High-Bandwidth Memory Stencil
  Accelerator for Weather Prediction Modeling},'' in \emph{FPL}, 2020.

\bibitem{seshadri.bookchapter17}
V.~Seshadri and O.~Mutlu, ``{Simple Operations in Memory to Reduce Data
  Movement},'' in \emph{Advances in Computers, Volume 106}, 2017.

\bibitem{diab2022high}
S.~Diab \emph{et~al.}, ``{High-throughput Pairwise Alignment with the Wavefront
  Algorithm using Processing-in-Memory},'' \emph{arXiv preprint
  arXiv:2204.02085}, 2022.

\bibitem{diab2022hicomb}
S.~Diab \emph{et~al.}, ``{High-throughput Pairwise Alignment with the Wavefront
  Algorithm using Processing-in-Memory},'' in \emph{HICOMB}, 2022.

\bibitem{fujiki2018memory}
D.~Fujiki \emph{et~al.}, ``{In-Memory Data Parallel Processor},'' in
  \emph{ASPLOS}, 2018.

\bibitem{zha2020hyper}
Y.~Zha and J.~Li, ``{Hyper-AP: Enhancing Associative Processing Through A
  Full-Stack Optimization},'' in \emph{ISCA}, 2020.

\bibitem{mutlu.imw13}
O.~Mutlu, ``{Memory Scaling: A Systems Architecture Perspective},'' \emph{IMW},
  2013.

\bibitem{mutlu.superfri15}
O.~Mutlu and L.~Subramanian, ``{Research Problems and Opportunities in Memory
  Systems},'' \emph{SUPERFRI}, 2014.

\bibitem{ahmed2019compiler}
H.~Ahmed \emph{et~al.}, ``{A Compiler for Automatic Selection of Suitable
  Processing-in-Memory Instructions},'' in \emph{DATE}, 2019.

\bibitem{jain2018computing}
S.~Jain \emph{et~al.}, ``{Computing-in-Memory with Spintronics},'' in
  \emph{DATE}, 2018.

\bibitem{ghiasi2022genstore}
N.~M. Ghiasi \emph{et~al.}, ``{GenStore: A High-Performance and
  Energy-Efficient In-Storage Computing System for Genome Sequence Analysis},''
  in \emph{ASPLOS}, 2022.

\bibitem{deoliveira2021IEEE}
G.~F. Oliveira \emph{et~al.}, ``{DAMOV: A New Methodology and Benchmark Suite
  for Evaluating Data Movement Bottlenecks},'' \emph{IEEE Access}, 2021.

\bibitem{deoliveira2021}
G.~F. Oliveira \emph{et~al.}, ``{DAMOV: A New Methodology and Benchmark Suite
  for Evaluating Data Movement Bottlenecks},'' \emph{arXiv:2105.03725 [cs.AR]},
  2021.

\bibitem{cho2020mcdram}
S.~Cho \emph{et~al.}, ``{McDRAM v2: In-Dynamic Random Access Memory Systolic
  Array Accelerator to Address the Large Model Problem in Deep Neural Networks
  on the Edge},'' \emph{IEEE Access}, 2020.

\bibitem{shin2018mcdram}
H.~Shin \emph{et~al.}, ``{McDRAM: Low latency and energy-efficient matrix
  computations in DRAM},'' \emph{IEEE TCADICS}, 2018.

\bibitem{gu2020ipim}
P.~Gu \emph{et~al.}, ``{iPIM: Programmable In-Memory Image Processing
  Accelerator using Near-Bank Architecture},'' in \emph{ISCA}, 2020.

\bibitem{lavenier2020}
D.~{Lavenier} \emph{et~al.}, ``{Variant Calling Parallelization on
  Processor-in-Memory Architecture},'' in \emph{BIBM}, 2020.

\bibitem{Zois2018}
V.~Zois \emph{et~al.}, ``{Massively Parallel Skyline Computation for
  Processing-in-Memory Architectures},'' in \emph{PACT}, 2018.

\bibitem{weber2005current}
D.~Weber \emph{et~al.}, ``{Current and Future Challenges of DRAM
  Metallization},'' in \emph{IITC}, 2005.

\bibitem{peng2015design}
Y.~Peng \emph{et~al.}, ``{Design, Packaging, and Architectural Policy
  Co-optimization for DC Power Integrity in 3D DRAM},'' in \emph{DAC}, 2015.

\bibitem{devaux2019}
F.~{Devaux}, ``{The True Processing In Memory Accelerator},'' in \emph{Hot
  Chips}, 2019.

\bibitem{yuffe2011}
M.~Yuffe \emph{et~al.}, ``{A Fully Integrated Multi-CPU, GPU and Memory
  Controller 32nm processor},'' in \emph{ISSCC}, 2011.

\bibitem{christy2020}
R.~Christy \emph{et~al.}, ``{8.3 A 3GHz ARM Neoverse N1 CPU in 7nm FinFET for
  Infrastructure Applications},'' in \emph{ISSCC}, 2020.

\bibitem{singh2017}
T.~Singh \emph{et~al.}, ``{3.2 Zen: A Next-generation High-performance x86
  Core},'' in \emph{ISSCC}, 2017.

\bibitem{upmem}
UPMEM, ``{UPMEM Website},'' \url{https://www.upmem.com}, 2020.

\bibitem{upmem2018}
UPMEM, ``{Introduction to UPMEM PIM. Processing-in-memory (PIM) on DRAM
  Accelerator (White Paper)},'' 2018.

\bibitem{gomezluna2021benchmarking}
J.~G{\'o}mez-Luna \emph{et~al.}, ``{Benchmarking a New Paradigm: An
  Experimental Analysis of a Real Processing-in-Memory Architecture},''
  arXiv:2105.03814 [cs.AR], 2021.

\bibitem{gomezluna2022ieeeaccess}
J.~Gómez-Luna \emph{et~al.}, ``{Benchmarking a New Paradigm: Experimental
  Analysis and Characterization of a Real Processing-in-Memory System},''
  \emph{IEEE Access}, 2022.

\bibitem{gomezluna2021cut}
J.~G{\'o}mez-Luna \emph{et~al.}, ``{Benchmarking Memory-Centric Computing
  Systems: Analysis of Real Processing-In-Memory Hardware},'' in \emph{IGSC},
  2021.

\bibitem{kwon202125}
Y.-C. Kwon \emph{et~al.}, ``{25.4 A 20nm 6GB Function-In-Memory DRAM, Based on
  HBM2 with a 1.2 TFLOPS Programmable Computing Unit Using Bank-Level
  Parallelism, for Machine Learning Applications},'' in \emph{ISSCC}, 2021.

\bibitem{lee2021hardware}
S.~Lee \emph{et~al.}, ``{Hardware Architecture and Software Stack for PIM Based
  on Commercial DRAM Technology: Industrial Product},'' in \emph{ISCA}, 2021.

\bibitem{ke2021near}
L.~Ke \emph{et~al.}, ``{Near-Memory Processing in Action: Accelerating
  Personalized Recommendation with AxDIMM},'' \emph{IEEE Micro}, 2021.

\bibitem{jedec.hbm.spec}
{JEDEC}, ``{High Bandwidth Memory (HBM) DRAM},'' Standard No. JESD235, 2013.

\bibitem{lee.taco16}
D.~Lee \emph{et~al.}, ``{Simultaneous Multi-Layer Access: Improving 3D-Stacked
  Memory Bandwidth at Low Cost},'' \emph{TACO}, 2016.

\bibitem{lee2022isscc}
S.~Lee \emph{et~al.}, ``{A 1ynm 1.25V 8Gb, 16Gb/s/pin GDDR6-based
  Accelerator-in-Memory supporting 1TFLOPS MAC Operation and Various Activation
  Functions for Deep-Learning Applications},'' in \emph{ISSCC}, 2022.

\bibitem{niu2022isscc}
D.~Niu \emph{et~al.}, ``{184QPS/W 64Mb/mm2 3D Logic-to-DRAM Hybrid Bonding with
  Process-Near-Memory Engine for Recommendation System},'' in \emph{ISSCC},
  2022.

\bibitem{freedman2009statistical}
D.~A. Freedman, \emph{{Statistical Models: Theory and Practice}}, 2009.

\bibitem{yan2009linear}
X.~Yan and X.~Su, \emph{{Linear Regression Analysis: Theory and Computing}},
  2009.

\bibitem{hosmer2013applied}
D.~W. Hosmer~Jr \emph{et~al.}, \emph{{Applied Logistic Regression}}, 2013.

\bibitem{suthaharan2016decision}
S.~Suthaharan, ``{Decision Tree Learning},'' in \emph{Machine Learning Models
  and Algorithms for Big Data Classification}, 2016.

\bibitem{Lloyd82leastsquares}
S.~P. Lloyd, ``{Least Squares Quantization in PCM},'' \emph{IEEE Transactions
  on Information Theory}, 1982.

\bibitem{hwukirk2016.nn}
D.~B. Kirk \emph{et~al.}, \emph{{Programming Massively Parallel Processors, 3rd
  Edition, Chapter 16 - Application Case Study: Machine Learning}}.\hskip 1em
  plus 0.5em minus 0.4em\relax Morgan Kaufmann, 2017.

\bibitem{chetlur2014cudnn}
S.~Chetlur \emph{et~al.}, ``{cuDNN: Efficient Primitives for Deep Learning},''
  \emph{arXiv preprint arXiv:1410.0759}, 2014.

\bibitem{abadi2016tensorflow}
M.~Abadi \emph{et~al.}, ``{Tensorflow: A System for Large-scale Machine
  Learning},'' in \emph{OSDI}, 2016.

\bibitem{runai_gpu}
Run:AI, ``{Best GPU for Deep Learning},''
  \url{https://www.run.ai/guides/gpu-deep-learning/best-gpu-for-deep-learning/},
  2021.

\bibitem{jouppi2017datacenter}
N.~P. Jouppi \emph{et~al.}, ``{In-Datacenter Performance Analysis of a Tensor
  Processing Unit},'' in \emph{ISCA}, 2017.

\bibitem{jouppi2021ten}
N.~P. Jouppi \emph{et~al.}, ``{Ten Lessons from Three Generations Shaped
  Google’s TPUv4i: Industrial Product},'' in \emph{ISCA}, 2021.

\bibitem{upmem-guide}
UPMEM, ``{UPMEM User Manual. Version 2021.3.0},'' 2021.

\bibitem{upmem-sdk}
UPMEM, ``{UPMEM Software Development Kit (SDK).}'' \url{https://sdk.upmem.com},
  2021.

\bibitem{zmora2021quantization}
N.~Zmora \emph{et~al.}, ``{Achieving FP32 Accuracy for INT8 Inference Using
  Quantization Aware Training with NVIDIA TensorRT},''
  \url{https://developer.nvidia.com/blog/achieving-fp32-accuracy-for-int8-inference-using-quantization-aware-training-with-tensorrt/}.

\bibitem{gholamisurvey}
A.~Gholami \emph{et~al.}, ``{A Survey of Quantization Methods for Efficient
  Neural Network Inference},'' in \emph{Low-Power Computer Vision}.

\bibitem{hopper}
{NVIDIA}, ``{NVIDIA H100 Tensor Core GPU Architecture. White Paper},''
  \url{https://nvdam.widen.net/s/9bz6dw7dqr/gtc22-whitepaper-hopper}, 2022.

\bibitem{han1995influence}
J.~Han and C.~Moraga, ``{The Influence of the Sigmoid Function Parameters on
  the Speed of Backpropagation Learning},'' in \emph{IWANN}, 1995.

\bibitem{deng2019lacc}
Q.~Deng \emph{et~al.}, ``{LAcc: Exploiting Lookup Table-based Fast and Accurate
  Vector Multiplication in DRAM-based CNN Accelerator},'' in \emph{DAC}, 2019.

\bibitem{gao2016draf}
M.~Gao \emph{et~al.}, ``{DRAF: A Low-power DRAM-based Reconfigurable
  Acceleration Fabric},'' in \emph{ISCA}, 2016.

\bibitem{weisstein2004taylor}
E.~W. Weisstein, ``{Taylor Series},''
  \url{https://mathworld.wolfram.com/TaylorSeries.html}, 2004.

\bibitem{gomez2022experimental}
J.~G{\'o}mez-Luna \emph{et~al.}, ``{An Experimental Evaluation of Machine
  Learning Training on a Real Processing-in-Memory System},'' \emph{arXiv
  preprint arXiv:2207.07886}, 2022.

\end{thebibliography}
